\documentclass[12pt,aps,eqsecnum,nofootinbib,floatfix]{revtex4}
\UseRawInputEncoding
\usepackage{bbding}
\usepackage{pifont}
\usepackage{subfigure}
\usepackage{epsfig}
\usepackage{array}
\usepackage{amsmath}
\usepackage{amsfonts}
\usepackage{amssymb}
\usepackage{color}
\usepackage{graphicx}
\usepackage{mathrsfs}
\usepackage{multirow}
\usepackage{subfigure}
\usepackage[colorlinks,citecolor=blue,linkcolor=blue,urlcolor=blue,hyperindex,dvipdfm]{hyperref}

\def\be{\begin{equation}}
\def\ee{\end{equation}}
\def\bea{\begin{eqnarray}}
\def\eea{\end{eqnarray}}

\newcommand{\omits}[1]{}

\begin{document}

\title{Thermodynamics of the Reissner-Nordstr\"om-de Sitter Spacetime with Quintessence}
\author{Yang Zhang}
\author{Yu bo Ma}
\author{Yun Zhi Du}
\email{duyzh13@lzu.edu.cn}
\author{Huai Fan Li}
\email{huaifan999@163.com}
\author{Li Chun Zhang}
\affiliation{ Institute of Theoretical Physics, Shanxi Datong University, Datong,
037009, China}

\begin{abstract}
For Anti-de Sitte (AdS) black holes, the isochoric heat capacity of system is vanished,
while the isobaric heat capacity is not. However, this situation does not hold on for de Sitter (dS) black holes. In this work, by introducing the interaction between the black hole horizon and the cosmological horizon of the Reissner-Nordstr\"om-de Sitter (RNdS) spacetime with quintessence, we discuss the phase transition of this system. The results show that the spacetime not only has the similar phase transition behavior to that of Van der Waals (VdW) system, and the non-vanishing isochoric heat capacity fulfills the whole thermodynamics system. Through the discussion of the entropic force between two horizons, we find out the role of entropic force in the evolution of spacetime. In addition, we also study the influence of various
parameters on the phase transition and entropic force, which will provide a new
method for exploring the interaction among black hole molecules from a micro
perspective.

\end{abstract}

\maketitle

\section{Introduction}

As we well known, a black hole is a celestial body whose gravity is so strong that even light cannot escape. Recently, the cooperation groups of LIGO and Virgo have repeatedly observed gravitational wave signals from double black hole merging events, and the cooperation group of EHT photographed the shadow of the supermassive black hole in the center of the $M87$ elliptical galaxy. These are observational evidence of the strong gravitational effect of black holes. In fact, a black hole is not only a strong gravitational system, but also a thermodynamic system. As early as $1970$s, physicists such as Hawking and Bekenstein have established the four laws of black hole thermodynamics.\cite{Hawking-1975,Bekenstein-1973,Bardeen-1973,Hawking-1983}

In order to explore the microstructure and evolution of black holes, people regard the cosmological constant of $n$-dimensional AdS spacetime as the state parameter of the black hole thermodynamic system - pressure
\begin{equation}
P=\frac{{n(n-1)}}{{16\pi{l^2}}},~\Lambda=-\frac{{n(n-1)}}{{2{l^2}}},
\end{equation}
whose conjugate thermodynamic quantity is the volume
\begin{equation}
V={\left({\frac{{\partial M}}{{\partial P}}}\right)_{S,{Q_i},{J_k}}}.
\end{equation}
Based on this issue, a series of studies on the thermodynamic behavior of AdS black holes have been done in the extended phase space \cite{David-2012,rong-2013,Sharmila-2012,Antonia-2014, Joy-2020,cai-2016,jia-2015,Hendi-2017,Dehghani-2020,Hadi-2020,Chabab-2019,
Daniela-2020,Sajadi-2019,Hendi-2019,Guo-2021,Hendi-2021,xu-2021,Amin-2020,Wen-2020}. Firstly, people found that the charged AdS black hole has a behavior similar to Van de Waals phase transition when choosing the quantities $P$ and $V$ as the independent dual variables. Subsequently, it has been proved in Ref. \cite{Amin-201777} that a charged AdS black hole possesses the Van der Waals (VdW) phase transition behavior by choosing $Q^2-\Psi$ as the independent dual variables. These developments of AdS black hole thermodynamic properties will not only help us understand the nature of black holes, but also help us to explore the microstructure of black holes \cite{wei-2019,wei-2020,zou-2020,wei-202011,miao-2018,
miao-2019,guo-2020,guo-2019,mann-2021,Volovik-2021}.

In addition, since our universe during the inflationary period was a quasi-de Sitter spacetime, that drives our attention to a dS spacetime. Considering the cosmological constant in a dS spacetime as the dark energy, our accelerated expanding universe will evolve into a new de Sitter phase in the future \cite{cai-2002}. Therefore, we should pay more attention on the relation between the classical, quantum, and thermodynamic properties of dS spacetime. While there are merely works on the thermodynamic behavior of black holes in a dS spacetime\cite{Yuichi-2006, Miho-2009,Saoussen-2019,David-2016, Simovic-2019,Sumarna-2020,Simovic-202020,Chabab-202020,Brian-2013,Bhattacharya-2016,James-2016, Kanti-2017,Romans-1992,zhang-2016,zhang-2019}. The related thermodynamic properties of dS black holes will help us understand the relation between gravity and the conformal field theory. However, the investigation of the black hole thermodynamic properties in a dS spacetime is a complicated problem, since the absence of an everywhere timelike Killing vector outside the black hole horizon leads to an unclear definition of asymptotic mass. Furthermore, in a dS spacetime the black hole and cosmological horizons yield two distinct temperatures, which suggests that the system is in a non-equilibrium state.

The authors in Refs. \cite{ma-2020,guoxiong-2020} had shown that when the interaction between the black hole and cosmological horizons of the Reissner-Nordstr\"om-de Sitter black hole surrounded by quintessence (RN-dSQ) is considered, the entropy is the summation of the corresponding entropy for two horizons, and together with their interaction term, i.e.
\begin{equation}
S = {S_c}(1 + {x^2} + f(x)),
\end{equation}
where $S_c$ is the entropy of cosmological horizon and $x$ is ratio of positions of two horizons and the interaction term reads
\begin{equation}
f(x)=\frac{8}{5}{(1-{x^3})^{2/3}}-\frac{{2(4-5{x^3}-{x^5})}}{{5(1-{x^3})}}.
\end{equation}
The corresponding effective temperature satisfies ${T_{eff}}={T_+}={T_c}$, when the two horizon temperatures are equal. This result fulfills the requirements of ordinary thermodynamic systems. Therefore, the effective thermodynamic quantities of a dS spacetime are more general after considering the interaction between two horizons.

%In this paper, we firstly discuss the conditions for the existence of black hole and cosmological horizons in RN-dSQ spacetime, especially the influence of spacetime parameters on the existence of two horizons. Then the minimum value of the position ratio of two horizons is found through analysis, namely, the range of $x$ is presented.

In Refs. \cite{ma-2020,guoxiong-2020} the authors assumed that there exists the interplay between the black hole and cosmological horizons in the RN-dSQ spacetime. And on this basis, they gave the effective thermodynamic quantities of spacetime, further discussed the thermodynamic properties of this system. Results shown that this system also has a continuous phase transition similar to the Van de Waals system or a charged AdS black hole, and parameters of this system influence the phase transition.
%By means of identifying the spherically symmetric AdS black hole with a thermodynamic system, and using the research methods of ordinary thermodynamic systems for reference, people obtained the thermodynamic characteristics of the spherically symmetric AdS black hole.
However, a vanishing isochoric heat capacity in this case is unacceptable, which has aroused people's attention
\cite{Jack-2020,Clifford-2020}. When the effective thermodynamic quantities are used to describe the thermodynamic properties of spacetime, the isochoric heat capacity will not vanish. It is shown that, under the consideration of interaction between two horizons, the method of the effective thermodynamic quantities to describe the thermodynamic properties of de Sitter spacetime has a more universal physical meaning. Furthermore, from the total entropy between two horizons in a RN-dSQ spacetime, the entropic force in the evolution of a dS spacetime plays a important role, which presents a new approach for studying the interaction of microscopic particles inside black holes and simulating the evolution of our accelerated universe.

This article is organized as follows. In Sec. \ref{2}, we discuss the conditions for the existence of black hole and cosmological horizons in RN-dSQ spacetime, as well as the influence of spacetime parameters, and give the range of the position ratio $x$ of two horizons. In Sec. \ref{3}, we present the effective thermodynamic quantities of RN-dSQ spacetime, and analyse the relation between the effective temperature $T_{eff}$ and the position ratio $x$ of two horizons. The merely consistency between the effective temperature and black hole horizon temperature reminds us that when studying the spacetime thermodynamic properties of RN-dSQ, replacing the effective temperature $T_{eff}$ with the black hole horizon radiation temperature $T$ can simplify
the problem. In Sec. \ref{4}, we extend the method of the phase transition for Van de Waals systems or charged AdS black holes, and use it to study the RN-dSQ spacetime, and obtain the phase transition properties of the RN-dSQ spacetime. %According to Ehrenfest's classification for phase transitions, it is found that RN-dSQ space-time has continuous and first-order phase transitions, and the influence of spacetime parameters on phase transitions is analyzed, too.
In Sec. \ref{5}, we explore the entropic force between two horizons of the RN-dSQ spacetime, and obtain the function of entropic force with respect to position ratio $x$. It is found that the law of entropic force changing with $x$ is very similar to the
Lennard-Jones force between two microscopic particles. This phenomenon provides a new way for us to further study the interaction of black hole molecules. The Sec. \ref{6} is the summary.

\section{Reissner-Nordstr\"om-de Sitter black hole surrounded by quintessence}
\label{2}

When solving the Einstein-Maxwell field equation with the cosmological constant, in $2003$ Kiselev considered the quintessence dark energy
\begin{eqnarray}
T_t^t&=&T_r^r=\rho_q+\frac{\alpha}{{r^2}},\\
T_\theta^\theta&=&T_\varphi^\varphi=-\frac{1}{2}{\rho_q}(3\omega+1),
\end{eqnarray}
where $\rho_q=-\frac{{3\alpha \omega }}{{2{r^{3(\omega+1)}}}}$ is the density of quintessence field, $\omega$ ($-1<\omega<-1/3$) is the quintessence dark energy barotropic index, $\alpha$ is the normalization factor associated with the quintessence density of dark energy and is always positive. On this issue, the static spherically symmetric solution of Einstein-Maxwell field equation for a Reissner-Nordstr\"om-dS (RN-dS) black hole in quintessence matters had been given by Refs. \cite{Kiselev-2003,Hang-2017,Chabab-201818,Huang-2021,Hong-2019}
\begin{equation}
d{s^2} = - g(r)d{t^2} + {g^{ - 1}}d{r^2} + {r^2}d\Omega _2^2
\end{equation}
with the metric function
\begin{equation}
g(r)=1-\frac{{2M}}{r}+\frac{{Q^2}}{{r^2}}-\frac{{r^2}}{{l^2}}-\alpha{r^{-3\omega-1}},
\end{equation}
here $M$ and $Q$ represent the black hole mass and charge, $l$ is the curvature radius of dS space.
\begin{figure}[tbp]
\centering
{\includegraphics[width=0.3\columnwidth,height=1.5in]{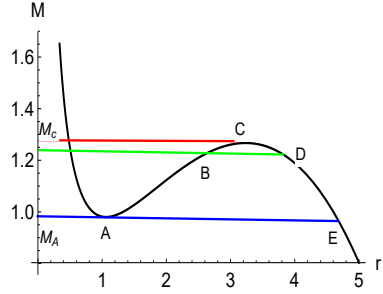}}
\caption{(color online).{$M - r$ diagram for $Q=1$, $\alpha=0.01$, $\omega=-2/3$, $l^2=37.386$.}}\label{Fig1}
\end{figure}
\begin{figure}[tbp]
\centering
\subfigure[~$\omega=-2/3,~\alpha=0.01$]{
{\includegraphics[width=0.3\columnwidth,height=1.5in]{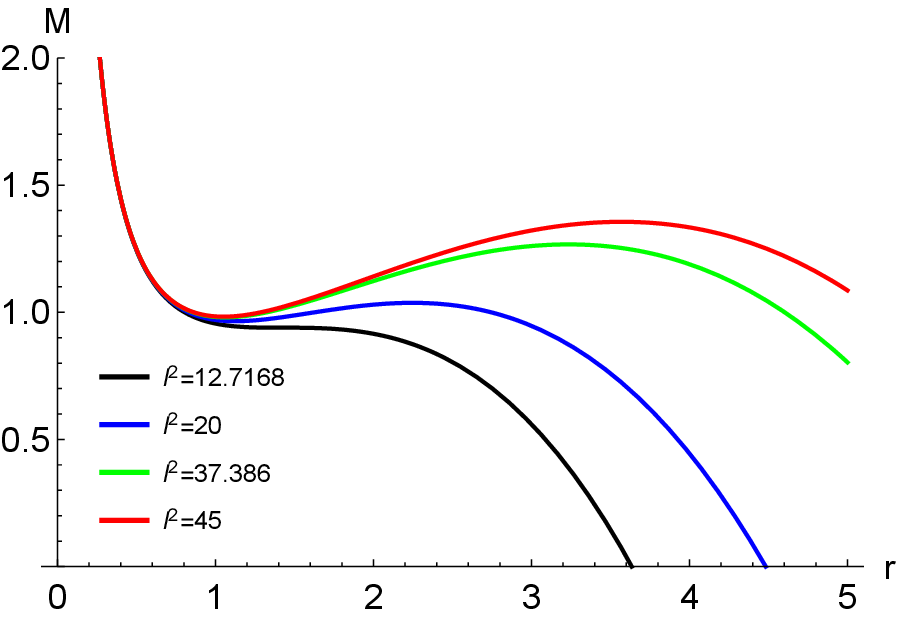}}\label{Fig2.1}}~~
\subfigure[~$\omega=-2/3,~l^{2}=37.386$]{
{\includegraphics[width=0.3\columnwidth,height=1.5in]{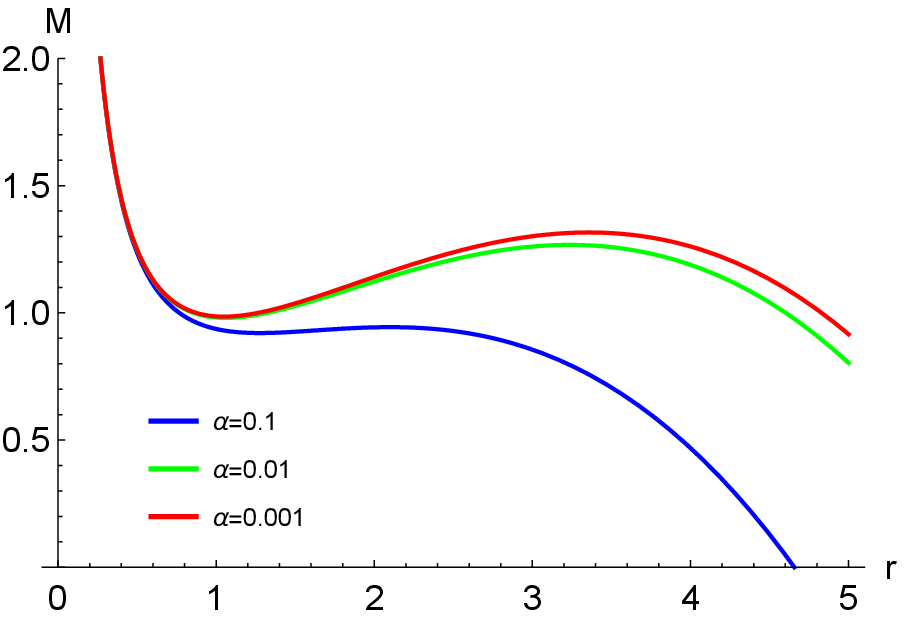}}\label{Fig2.2}}~~
\subfigure[~$l^{2}=37.386,~\alpha=0.01$]{
{\includegraphics[width=0.3\columnwidth,height=1.5in]{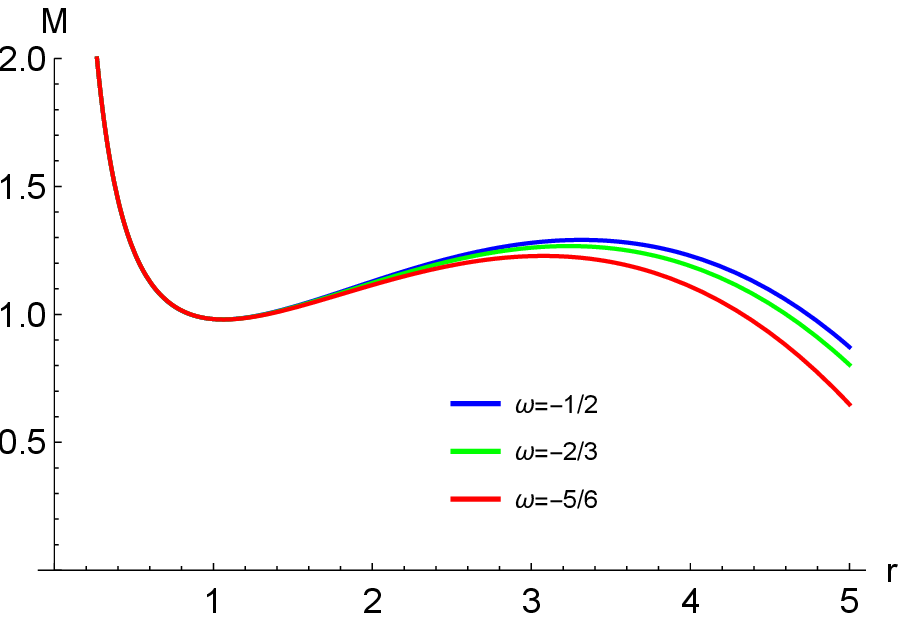}}\label{Fig2.3}}
\caption{(color online). $M - r$ curves with different values of  $l^{2}$, $\alpha$, and $\omega$ with $Q = 1$.}
\end{figure}

Now, let us analyze the existence of black hole and cosmological horizons, which are determined by the following equation
\begin{equation}
1 - \frac{{2M}}{r} + \frac{{Q^2}}{{r^2}} - \frac{{r^2}}{{l^2}} -
\alpha {r^{ - 3\omega - 1}} = 0.\label{2-7}
\end{equation}
Solving Eq. (\ref{2-7}), one can obtain
\begin{equation}
M = \frac{r}{2}\left( {1 + \frac{{Q^2}}{{r^2}} - \frac{{r^2}}{{l^2}} -
\alpha {r^{ - 3\omega - 1}}} \right). \label{2-8}
\end{equation}
By choosing $Q=1$, $\omega=-2/3$, $l^2=37.386$, $\alpha=0.01$, the $M-r$  curve is shown in Fig. \ref{Fig1}. It can be seen from the $M-r$ graph that when choose $M_A< M<M_C$, the RN-dSQ spacetime have the black hole interior horizon $r_-$, black hole exterior horizon $r_+$ (i.e., black hole horizon in the following), and the cosmological horizon $r_c$. At the point $C$, the black hole horizon coincides with the cosmological one, and this point also corresponds to the maximum energy of the black hole $M=M_C$. At the point $A$, the interior and exterior horizons of black hole coincide, and it is also the lowest limit of the smallest black
hole $M=M_A$. Black hole does not exist in RN-dSQ spacetime when $M>M_C$ or $M<M_A$ \cite{CaoH-2018}. In order to study the
thermodynamic properties of a dS spacetime endowed with these horizons, we should take $M_A\leq M\leq M_C$, and the values of $M_A$ and $M_C$ depend on the spacetime parameters.

Now we analyze the influence of parameters on $M-r$ curves, and the existence condition of the black hole and cosmological horizons. The Equation (\ref{2-8}) implies that when the parameters are chosen as $Q=1,~\omega=-2/3,~\alpha=0.01$, to ensure the existence of two horizons in RN-dSQ spacetime the minimum value of the cosmological constant is $l^2=12.7168$. Namely, there exist both the black hole and cosmological horizons in RN-dSQ spacetime as $l^2>12.7168$. On the other hand, the maximum value of $\alpha $ is $0.1$ under the same set of parameters. That is, when normalization factor associated with the quintessence density of dark energy $\alpha<0.1$, both the black
hole and cosmological horizons can survive. From Fig. \ref{Fig2.3} we can see that the value of $\omega$ has no effect on the existence of two horizons. Accordingly, two parameters which have effect on the existence of two horizons are the cosmological constant $l^2$ and the normalization factor associated with the quintessence density of dark energy $\alpha$.

Fig. \ref{Fig1} shows that when $M_{A}\leq M\leq M_{C},$ there exists the black hole horizon $r_{+}$ in RN-dSQ spacetime, which corresponds to the curve $A-B-C$. And the cosmological horizon $r_{c}$ corresponds to the curve $C-D-E$. And the points $A$ and $C$ correspond to the minimum and maximum energies of spacetime endow with two horizons, respectively. The cases of $M<M_{A}$ and $M>M_{C}$ are not within our consideration, because these two horizons cannot exist together in a
RN-dSQ spacetime. When the parameters $Q,~\omega,~\alpha$, from eq. (\ref{2-8}) are given, we can obtain the relations between the minimum value of cosmological constant ($l_{min}^2$) and horizon radius under which two horizons can both exist:
\begin{equation}
2{Q^{2}}-\frac{{6{r^{4}}}}{{{l_{min}^{2}}}}-3\omega (3\omega +1)\alpha {%
r^{-3\omega +1}}=0,{r^{2}}-{Q^{2}}-\frac{{3{r^{4}}}}{{{l_{min}^{2}}}}+3\omega
\alpha {r^{-3\omega +1}}=0.
\end{equation}
For different values of $Q,~\omega,~\alpha$, by solving above equations we present the minimum values of the cosmological constant in Table. \ref{TTT}.
\begin{table*}
\begin{center}
\caption[]{\it Minimum values of cosmological constant $l_{\min }^2$ for given
other parameters}\label{TTT}
\begin{tabular}{|c|c|c|c|c|c|c|} \hline
\cline{1-4}
   $\omega  = - 2/3$ &  $\alpha  = 0.1$   & $\alpha  = 0.01$    & $\alpha  = 0.001$   \\
        \hline
          $l_{\min }^2$     &26.3171    &12.7168    & 12.0682\\
        \hline
         $\alpha  = 0.01$   & $\omega  = - 1/2$   & $\omega  = - 2/3$   & $\omega  = - 5/6$\\
        \hline
              $l_{\min }^2$   & 12.4416    & 12.7168    & 13.1001\\
        \hline
\end{tabular}
\end{center}
\end{table*}

In addition, from Eq. (\ref{2-7}) we have
\begin{equation}
{r_{+}}\frac{{(1-x)}}{x}-\frac{{{Q^{2}}(1-x)}}{{{r_{+}}}}-\frac{{r_{+}^{3}(1-%
{x^{3}})}}{{{l^{2}}{x^{3}}}}-\alpha r_{+}^{^{-3\omega }}({x^{3\omega }}-1)=0,\label{2-10}
\end{equation}%
where $x\equiv{r_{+}}/{r_{c}}$ is ratio between the black hole horizon $r_{+}$ and cosmological horizon $r_{c}$. At the points $A$ and $C$, there exists the following equation
\begin{equation}
\frac{1}{2}-\frac{{{Q^{2}}}}{{2{r_+^{2}}}}-%
\frac{{3{r_+^{2}}}}{{2{l^{2}}}}+\frac{{3\omega \alpha }}{2}{r_+^{-3\omega -1}}=0. \label{2-11}
\end{equation}%
For simplification, we choose
\begin{equation}
\frac{{3\alpha }}{{r_{{+}}^{1+3\omega }}}=\frac{{3\alpha (4\pi r_{{+}}^{2})}}{{%
4\pi r_{{+}}^{3+3\omega }}}=-\frac{{{A_{{+}}}{\rho _{{+}}}}}{{2\pi \omega }}%
=-\frac{{{\beta _{{+}}}}}{{2\pi \omega }},
\end{equation}%
where $\beta_+$ is the quintessence field on the black hole horizon. Then Eq.
(\ref{2-11}) can be rewritten as
\begin{equation}
\frac{{{\beta _{+}}}}{{2\pi }}=\left( {1-\frac{{{Q^{2}}}}{{{r_+^{2}}}}-\frac{{%
3r_{+}^{2}}}{{{l^{2}}}}}\right),\label{2-13}
\end{equation}%
By combining Eq. (\ref{2-10}) and Eq. (\ref{2-11}), $r_{+}$ can be solved
\begin{equation}
r_{+}^{2}=\frac{{-b+\sqrt{{b^{2}}-4ac}}}{{2a}},\label{2-14}
\end{equation}%
where
\begin{equation}
a=-\frac{1}{{{l^{2}}}}\left( {\frac{{(1-{x^{3}})}}{{{x^{3}}}}+\frac{{{%
x^{3\omega }}}}{\omega }(1-{x^{-3\omega }})}\right),
\end{equation}%
\begin{equation}
b=\frac{{(1-x)}}{x}+\frac{{{x^{3\omega }}(1-{x^{-3\omega }})}}{{3\omega }},
c=-{Q^{2}}\left( {(1-x)+\frac{{{x^{3\omega }}(1-{x^{-3\omega }})}}{{3\omega
}}}\right) .
\end{equation}%
Using the definition $x_{min}=r_{A}/r_{E}$ shown in Fig. \ref{Fig1}, we can plot
the $x_{min}-l^{2}$ curves with respect to variant parameters by substituting Eq.
(\ref{2-14}) into Eq. (\ref{2-13}). And the numerical results of $x_{min}$ for
different parameters are shown in Table. \ref{Tab2}. The results show that the
minimum value $x_{min}$ is merely independent of $\beta _+$ and $l^2$, but it
is more sensitive to $\omega$, conversely.
\begin{table*}
\begin{center}
\caption[]{\it Minimum values of the ratio between two horizon $x_{\min }^2$ for
given other parameters}\label{Tab2}
\begin{tabular}{|c|c|c|c|c|c|c|} \hline
\cline{1-4}
   $\omega  = - 2/3$,  $l_{\min }^2=37.386$  &  ${\beta _{\rm{ +}}} = 0.0001$   & ${\beta _{\rm{ + }}} = 0.0005$    & ${\beta _{\rm{ + }}} = 0.0008$   \\
        \hline
          ${x_{\min }}$     & 0.431763    & 0.431753    & 0.431746\\
        \hline
          ${\beta _{\rm{ + }}} = 0.0005$, $l_{\min }^2=37.386$  & $\omega  = - 1/2$   & $\omega  = - 2/3$   & $\omega  = - 5/6$\\
        \hline
              $x_{\min }^2$   & 0.376711   & 0.431753    & 0.480403\\
        \hline
          ${\beta _{\rm{ + }}} = 0.0005$,  $\omega  = - 2/3$  & $l^2=20$   &  $l^2=37.386$   &  $l^2=45$\\
        \hline
              $x_{\min }^2$   & 0.43374   & 0.431753    & 0.430163\\
                   \hline
\end{tabular}
\end{center}
\end{table*}

 \begin{figure}[tbp]
\centering
%\hspace*{-0.5cm}
\subfigure[$\beta_+=0.0005$]{{\includegraphics[width=0.3\columnwidth,height=1.5in]{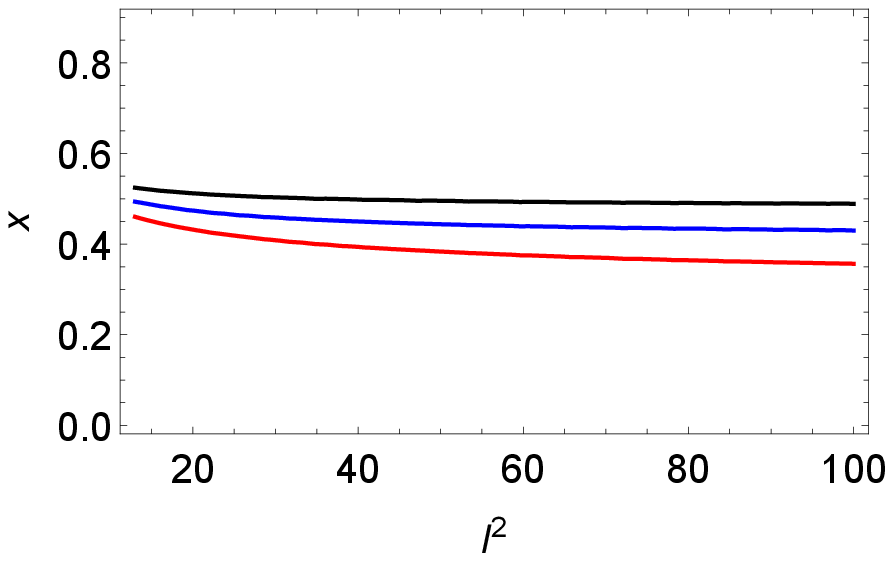}}\label{Fig3.1}}~~
\subfigure[$\omega=-1/2$]{{\includegraphics[width=0.3\columnwidth,height=1.5in]{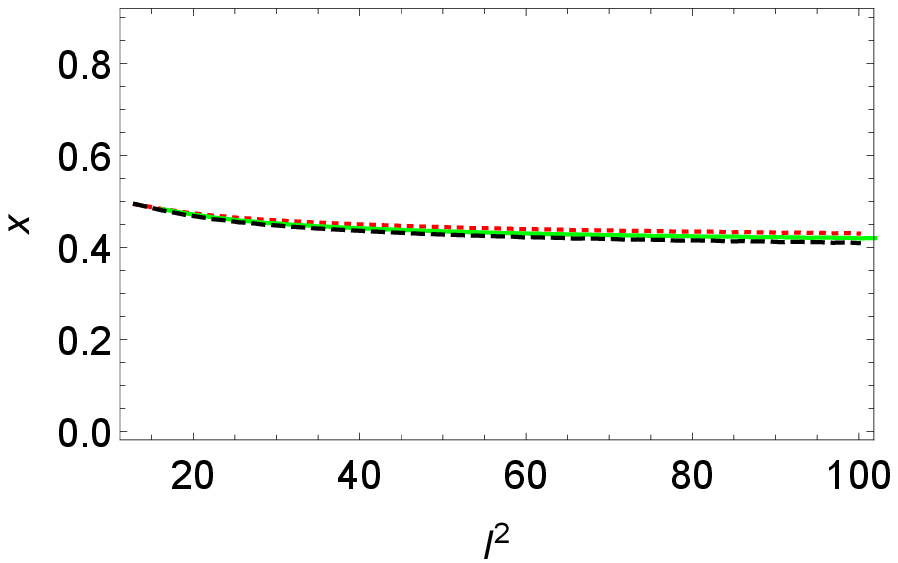}}\label{Fig3.2}}
\caption{(color online). Curves of ${x_{ min}} - l^2 $ with $Q = 1$. (a): the black solid line stands for $\omega=-1/2$, the blue solid line is for  $\omega=-2/3$, and the red solid line is for $\omega=-5/6$. (b): the red dotted line stands for $\beta _ + =0.0001$, the green solid line is for $\beta _ + =0.0005$, and the black dashed line is for $\beta _ + =0.0008$. }\label{Fig3}
\end{figure}
According to Eq. (\ref{2-13}), one has
\begin{equation}
r_{A,C}^2 = {l^2}\frac{{\left( {1 - \frac{{\beta _ + }}{{2\pi }}} \right)
\pm \sqrt {{{\left( {1 - \frac{{\beta _ + }}{{2\pi }}} \right)}^2} - 12{Q^2}/%
{l^2}} }}{6}.
\end{equation}
 \begin{figure}[tbp]
\centering
{\includegraphics[width=0.3\columnwidth,height=1.5in]{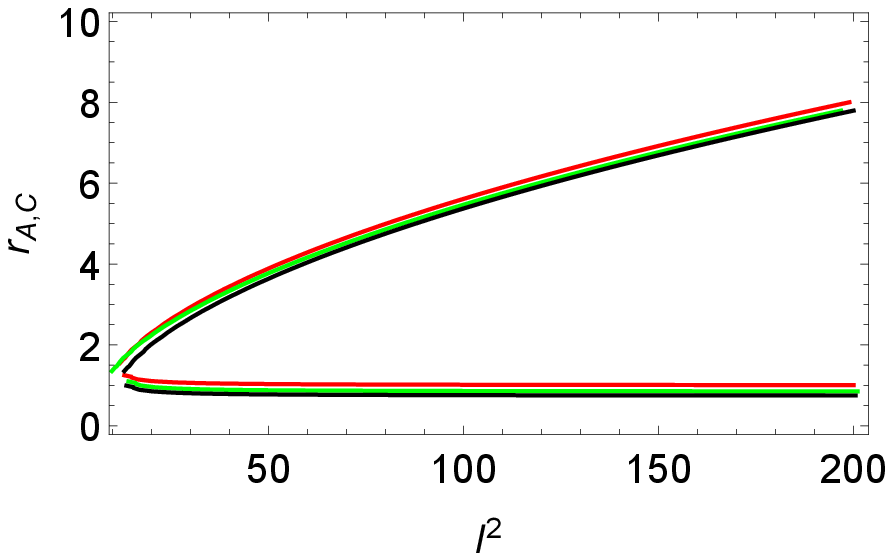}}
\caption{(color online). $r_{A,C}-l^2$ curves with $Q = 1$, $\omega=-1/2$.The red solid line stands for $\beta _ + =0.0001$, the green solid line for $\beta _ + =0.0005$, and the black solid line for $\beta _ + =0.0008$.}\label{Fig4}
\end{figure}
The above equation implies that the minimum radius $r_A$ and the maximum
radius $r_C$ of black hole horizon position in RN-dSQ spacetime are
independent of the value of $\omega$. Taking Q=1, $%
\beta_+=0.0001,0.0005,0.0008$, one can draw the curves of $%
r_{A,C}-l^2$, where one value of $l^2$ correspond to two values of $r$, i.e. $r_A$
and $r_C$. Therefore, the values of $M_A$ and $M_C$ depend on $l^2$ once $%
\beta_+$ and $Q$ are fixed. Fig. \ref{Fig4} shows that $r_C$ increases with the increase of $l^2$, while $r_A$
remains nearly unchanged, which is consistent with the fact that $x_{min}$
is independent on $l^2$.

To study the thermodynamic properties of RN-dSQ spacetime , we need to study
the thermodynamic properties of RN-dSQ spacetime endowed with the black hole
horizon and the cosmological horizon. While the existence of these two
horizons is determined by their position ratio $x$. Fig. \ref{Fig2.1} implies that the
range of $x$ is $x_{min}\leq x\leq1$ once the spacetime parameters are set, and $%
x_{min}$ is affected by parameters as shown in Figs. \ref{Fig3.1} and \ref{Fig3.2}.
Therefore, we adopt $x_{min}\leq x\leq1$ as the range of ratio $x$ to study
the thermodynamic properties of RN-dSQ spacetime .

\section{Effective thermodynamic quantities of RN-dSQ spacetime}
\label{3}
For the RN-dSQ spacetime, we focus on the space between the black hole and the cosmological horizon. Thus the volume of system reads \cite{Simovic-2019,Sumarna-2020,zhang-2016,zhang-2019}
\begin{equation}
V=\frac{{4\pi }}{3}\left( {r_{c}^{3}-r_{+}^{3}}\right) =\frac{{4\pi }}{{3{%
x^{3}}}}r_{+}^{3}(1-{x^{3}}).\label{3-2}
\end{equation}
When introducing the interaction between the two horizons, the total entropy of system is the sum of two individual horizon entropies and their interactive term, which is a function of positions of the two horizons
\begin{equation}
S=\pi r_{c}^{2}(1+{x^{2}}+f(x))=\pi r_{+}^{2}(1+{x^{2}}+f(x))/{x^{2}}.\label{3-3}
\end{equation}%
Here the undefined function $f(x)$ represents the extra contribution from the correlations of two horizons. Based on the first law of black hole thermodynamics \cite{zhang-2016,zhang-2019,ma-2020,guoxiong-2020}:
\begin{equation}
dM={T_{eff}}dS-{P_{eff}}dV+{\phi _{eff}}dQ\label{3-1}
\end{equation}
and Eqs. (\ref{3-2}), (\ref{3-3}), and (\ref{3-1}), the interactive function $f(x)$, effective temperature, and effective pressure of this system have the following forms \cite{zhang-2016,zhang-2019,Liu-2019}
\begin{eqnarray}
&&f(x)=\frac{8}{5}{\left( {1-{x^{3}}}\right) ^{2/3}}-\frac{{2\left( {4-5{x^{3}}%
-{x^{5}}}\right) }}{{5\left( {1-{x^{3}}}\right) }},\\
&&{T_{eff}}{f_{1}}(x)-\frac{{{f_{2}}(x)}}{{{r_{+}}}}+{Q^{2}}\frac{{{f_{3}}(x)}%
}{{r_{{+}}^{{3}}}}=0,\label{3-4}\\
&&{P_{eff}}{F_{1}}(x)r_{+}^{4}+{F_{2}}(x)r_{+}^{2}+{F_{3}}(x)=0\label{3-5}
\end{eqnarray}%
with
\begin{eqnarray}
{f_{1}}(x)&=&\frac{{4\pi (1+{x^{4}})}}{{1-x}},~~{f_{3}}(x)=(1+x+{x^{2}})(1+{x^{4}%
})-2{x^{3}},~~{F_{1}}(x)=\frac{{8\pi (1+{x^{4}})}}{{x(1-x)}},\nonumber\\
{f_{2}}(x)&=&(1+x)(1+{x^{3}})-2{x^{2}}-\frac{{{\beta _{{+}}}{x^{{3+3}\omega }}}%
}{{2\pi \omega (1-x)}}\left[ {1-{x^{-3\omega }}-\omega ({x^{3}}-{%
x^{-3-3\omega }})}\right],\nonumber\\
{F_{2}}(x) &=&\left({x(1+x)+\frac{{{\beta _{+}}{x^{4+3\omega }}(1-{x^{3-3\omega }}%
)}}{{2\pi (1-x)}}}\right)(x+f^{\prime}(x)/2) \nonumber\\
&\notag-&\left({\frac{{(1+2x)}}{{(1+x+{x^{2}})}}-\frac{{{\beta _{+}}{x^{3}}({x^{2\omega
}}-(2-\omega ){x^{5}}+(1-\omega ){x^{8}})}}{{2\pi \omega (1-{x^{3}})(1-x)}}}%
\right)(1+{x^{2}}+f(x)),\nonumber\\
{F_{3}}(x)&=&{Q^{2}}\left( {\frac{{(1+2x+3{x^{2}})}}{{(1+x+{x^{2}})}}(1+{x^{2}}%
+f(x))-(1+x+{x^{2}}+{x^{3}})(x+f^{\prime }(x)/2)}\right).\nonumber
\end{eqnarray}%
For the given effective temperature $T_{eff}$, from Eq. (\ref{3-4}) the inverse of black hole horizon radius satisfies
\begin{equation}
\frac{1}{r_+}=2\sqrt{\frac{{{f_{2}}(x)}}{{3{Q^{2}}{f_{3}}(x)}}}\cos (\theta +{%
120^{0}}).\label{3-9}
\end{equation}%
Once ${P_{eff}}$ is given, $r_{+}^{2}$ satisfies the following equation according to Eq. (3.5)
\begin{equation}
r_{+}^{2}=\frac{{-{F_{2}}(x)+\sqrt{F_{2}^{2}(x)-4{P_{eff}}{F_{1}}(x){F_{3}}%
(x)}}}{{2{P_{eff}}{F_{1}}(x)}}. \label{3-11}
\end{equation}%

\begin{figure}[tbp]
\centering
{\includegraphics[width=0.3\columnwidth,height=1.5in]{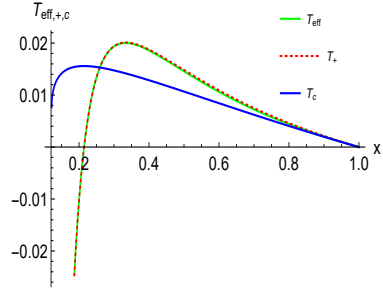}}
\caption{(color online). Lines show  ${T_{eff}}$, ${T_{+}}$ and ${T_{c}}$ as a function of $x$. }\label{Fig5}
\end{figure}

\begin{figure}[tbp]
\centering
\subfigure[$\omega=-2/3,~{l^{2}}=37.386$]{{\includegraphics[width=0.3\columnwidth,height=1.5in]{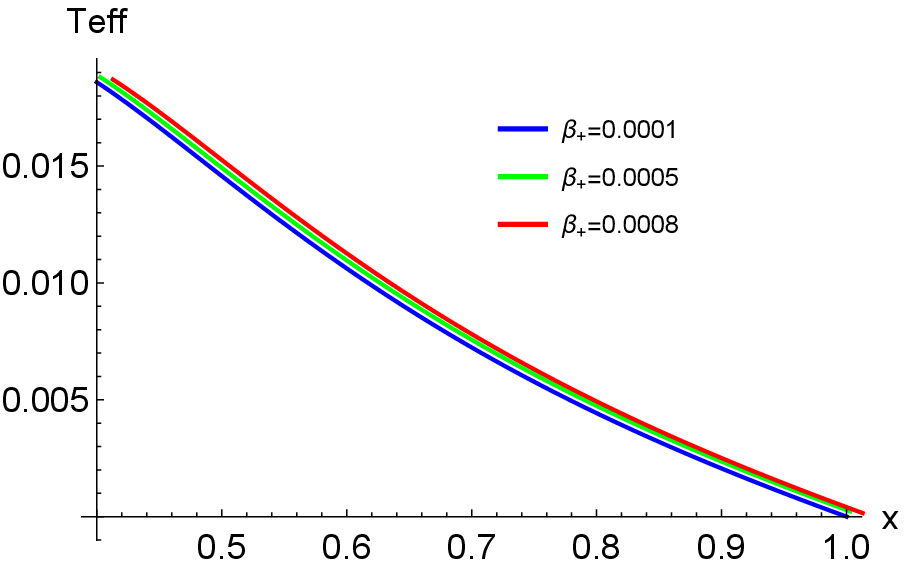}}}
\subfigure[$\omega=-2/3,~{\beta _{{+}}}=0.0005$]{{\includegraphics[width=0.3\columnwidth,height=1.5in]{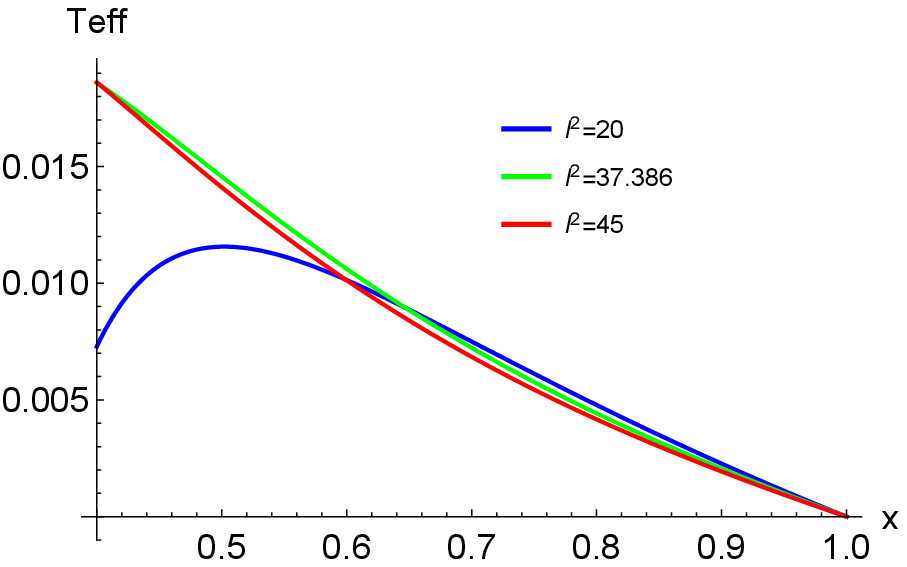}}}
\subfigure[${l^{2}}=37.386,~{\beta _{{+}}}=0.0005$]{{\includegraphics[width=0.3\columnwidth,height=1.5in]{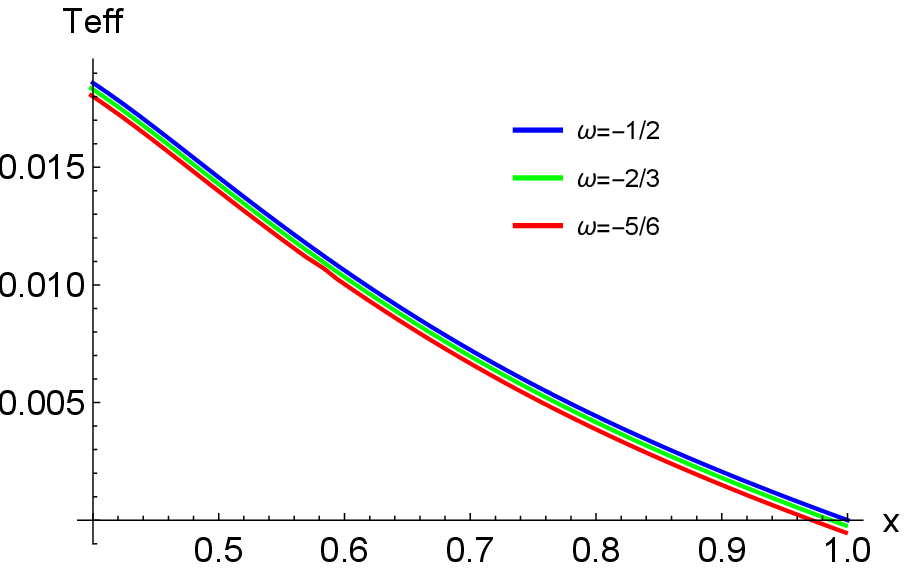}}}
\caption{(color online). ${T_{eff}}-x$ curves with $Q=1$.}\label{Fig6}
\end{figure}

In order to further understand the effective temperature of this system, we should compare it with the temperatures on two horizons. From Eq. (\ref{2-7}) we can obtain the Hawking temperatures of the black hole and the cosmological horizons as follows:
\begin{eqnarray}
{T_{+}}%=\frac{{g^{\prime }({r_{+}})}}{{4\pi }}=\frac{1}{{4\pi {r_{+}}}}%\left( {1-\frac{{{Q^{2}}}}{{r_{+}^{2}}}-\frac{{3r_{+}^{2}}}{{{l^{2}}}}%+3\omega \alpha r_{+}^{-3\omega -1}}\right)
&=&\frac{1}{{4\pi {r_{+}}}}\left( {%
1-\frac{{{Q^{2}}}}{{r_{+}^{2}}}-\frac{{3r_{+}^{2}}}{{{l^{2}}}}-\frac{{{\beta
_{+}}}}{{2\pi }}}\right), \label{3-12}\\
{T_{c}}%=-\frac{{g^{\prime }({r_{c}})}}{{4\pi }}=-\frac{1}{{4\pi {r_{c}}}}%\left( {1-\frac{{{Q^{2}}}}{{r_{c}^{2}}}-\frac{{3r_{c}^{2}}}{{{l^{2}}}}%+3\omega \alpha r_{c}^{-3\omega -1}}\right)
&=&-\frac{x}{{4\pi {r_{+}}}}\left(
{1-\frac{{{Q^{2}}{x^{2}}}}{{r_{+}^{2}}}-\frac{{3r_{+}^{2}}}{{{x^{2}}{l^{2}}}}%
-\frac{{{\beta _{+}}}}{{2\pi }}{x^{1+3\omega }}}\right)\label{3-13}
\end{eqnarray}%
%where
%\begin{equation}
%\frac{1}{{{l^{2}}}}=\frac{{{x^{2}}}}{{r_{+}^{2}(1+x+{x^{2}})}}-\frac{{{Q^{2}}%
%{x^{3}}}}{{r_{+}^{4}(1+x+{x^{2}})}}+\frac{{{\beta _{{+}}}(1-{x^{-3\omega }}){%
%x^{3+3\omega }}}}{{6\pi \omega (1-{x^{3}})}},
%\end{equation}%
with
\begin{equation}
r_{+}^{2}=\frac{{{x^{2}}}}{{(1+x+{x^{2}})}}\frac{{1\pm \sqrt{1-\frac{{4{Q^{2}%
}(1+x+{x^{2}})}}{x}\left( {\frac{1}{{{l^{2}}}}-\frac{{{\beta _{{+}}}(1-{%
x^{-3\omega }}){x^{3+3\omega }}}}{{6\pi \omega (1-{x^{3}})}}}\right) }}}{{%
2\left( {\frac{1}{{{l^{2}}}}-\frac{{{\beta _{{+}}}(1-{x^{-3\omega }}){%
x^{3+3\omega }}}}{{6\pi \omega (1-{x^{3}})}}}\right) }}.\label{3-15}
\end{equation}%
Substituting Eq. (\ref{3-15}) into Eq. (\ref{3-4}), (\ref{3-12}) and (\ref{3-13}), we plot the curves with the parameters $Q=1$, $\omega =-2/3$, ${\beta _{{+}}}=0.0005$, ${l^{2}=37.386}$ in Fig. \ref{Fig5}.
It is obviously that the trends of ${T_{eff}}$ and ${T_{+}}$ with respect to $x$ are same for the fixed parameters $Q$, $\omega $, ${\beta _{{+}}}$, and ${l^{2}}$. This implies the effective temperature can be replaced with the one on black hole horizon. In addition, the influence of parameters on the effective temperature is exhibited in Fig. \ref{Fig6}, from which we can see that the behavior of temperature as a function of $x$ is merely independent of $\omega $ and ${\beta _{{+}}}$ when $l^{2}$ is given. However, for the fixed parameters $Q$ ,$\omega $, and ${\beta _{{+}}}$, the behavior of ${T_{eff}}$ with respect to $x$ is only affected by $l^{2}$.

\section{phase transition of RN-dSQ spacetime}
\label{4}
\begin{figure}[tbp]
\centering
%\hspace*{-0.5cm}
\subfigure[$\omega=-1/2,~{\beta _{{+}}}=0.0005$]{{\includegraphics[width=0.3\columnwidth,height=1.5in]{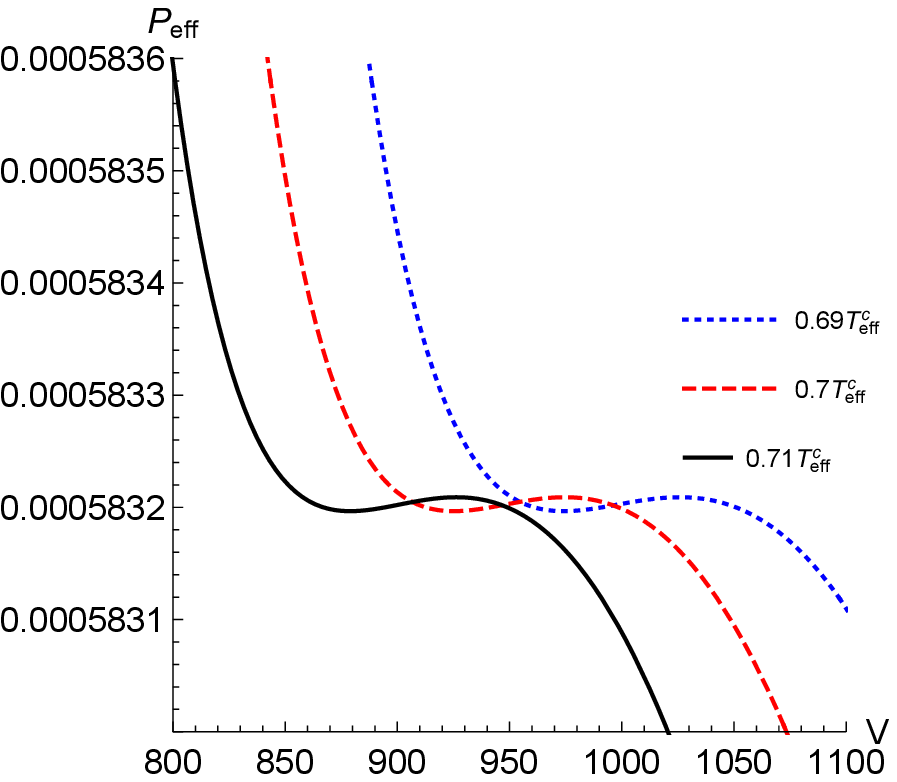}}\label{Fig7.1}}
\subfigure[${\beta _{{+}}}=0.0005,~~T_{eff}=T_{eff}^{c}$]{{\includegraphics[width=0.3\columnwidth,height=1.5in]{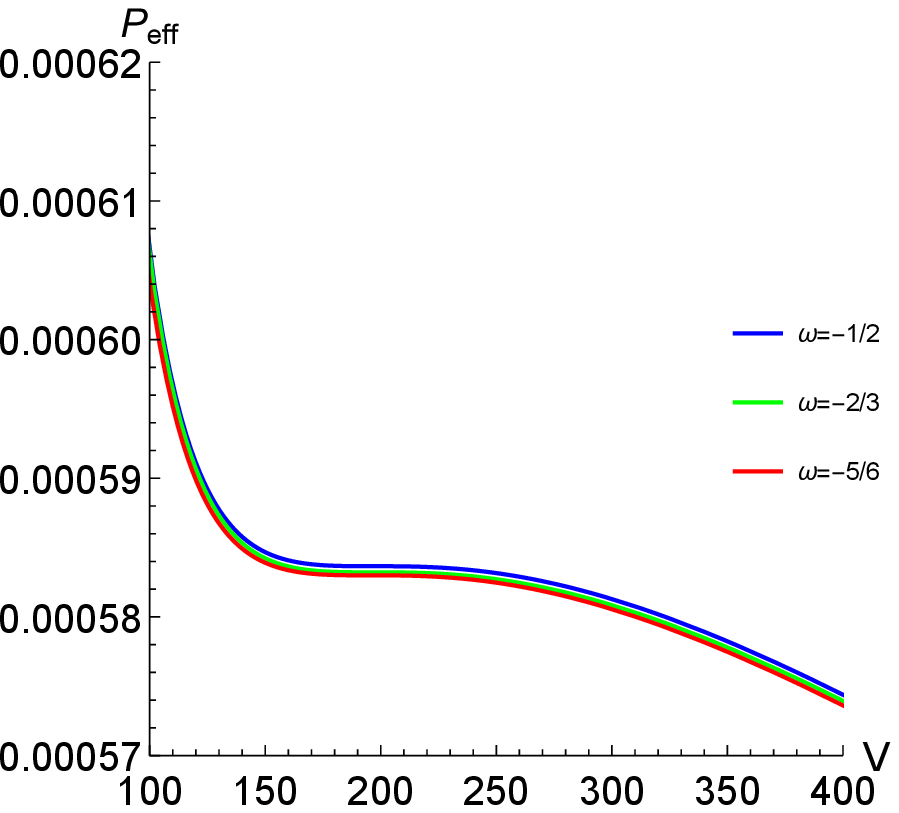}}\label{Fig7.2}}
\subfigure[$\omega=-1/2,~~T_{eff}=T_{eff}^{c}$]{{\includegraphics[width=0.3\columnwidth,height=1.5in]{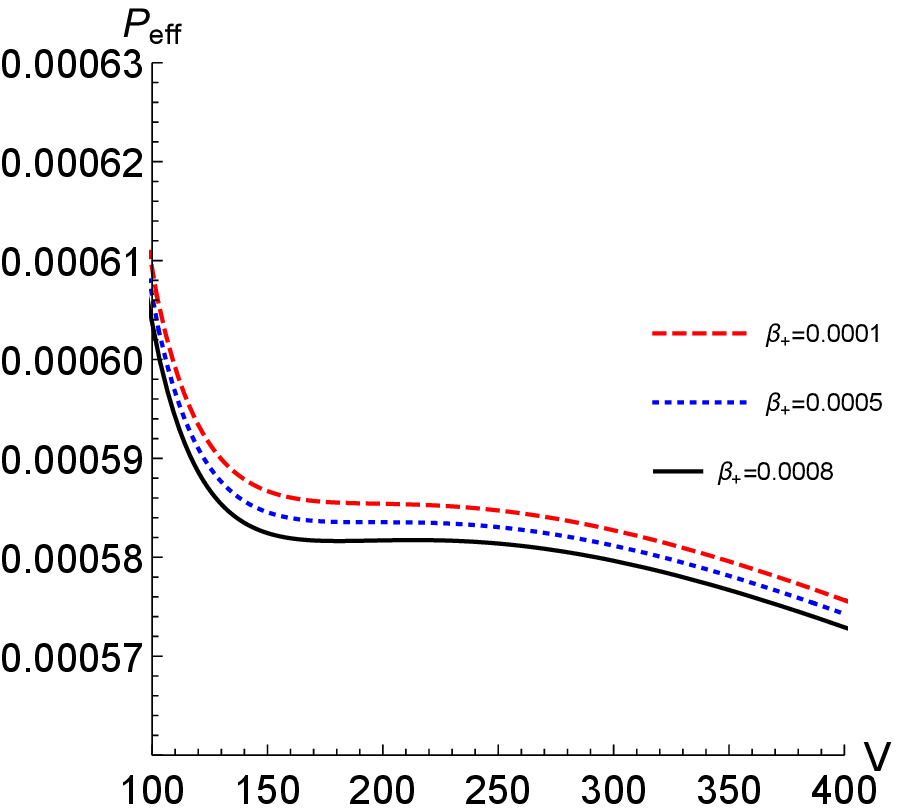}}\label{Fig7.3}}
%\label{c}}\newline
\caption{(color online). ${P_{eff}} - V$ curves for $Q = 1$.}\label{Fig7}
\end{figure}

In this section, we would like to study the phase transition in the
canonical ensemble. The Gibbs free energy of RN-dSQ spacetime reads
\begin{eqnarray}
G({r_{+}},x) &=&\frac{{{r_{+}}(1+x)}}{{2(1+x+{x^{2}})%
}}+\frac{{{Q^{2}}(1+x)(1+{x^{2}})}}{{2{r_{+}}(1+x+{x^{2}})}}+\frac{{{r_{+}}(1-x)D(x,\omega )}}{{6{x^{3}}(1+{x^{4}})}}(1-{x^{3}})\notag \\
&&+{\beta _{+}}{r_{+}}x\frac{{{x^{3\omega }}[(1+x)(1-{x^{3}})-2{x^{2}}%
]+(1+x)(1-{x^{2}})}}{{12\pi \omega (1+x+{x^{2}})(1-{x^{3}})}}  \notag \\
&&-\frac{{{r_{+}}(1-x)}}{{4{x^{2}}(1+{x^{4}})}}(1+{x^{2}}+{f_{0}}(x))\{{%
\left[ {(1+x)(1+{x^{3}})-2{x^{2}}}\right] }  \notag \\
&&-\frac{{{Q^{2}}}}{{r_{+}^{2}}}\left[ {(1+x+{x^{2}})(1+{x^{4}})-2{x^{3}}}%
\right] \notag \\
&&-\frac{{\beta _{+}{x^{3+3\omega }}}}{{2\pi \omega (1-x)}}{\left[ {1-{%
x^{-3\omega }}-\omega ({x^{3}}-{x^{-3-3\omega }})}\right] \}}.\label{4-1}
\end{eqnarray}
The critical point denotes a second-order phase transition which is determined by
the following equations
\begin{equation}
{\left( {\frac{{\partial {P_{eff}}}}{{\partial V}}}\right) _{{T_{eff}},Q,{%
\beta _{+}},\omega }}={\left( {\frac{{{\partial ^{2}}{P_{eff}}}}{{\partial {%
V^{2}}}}}\right) _{{T_{eff}},Q,{\beta _{+}},\omega }}=0.\label{4-2}
\end{equation}%
According to Eq. (\ref{4-2}), when choosing $Q=1$, ${\beta _{{+}}}=0.0005$, $\omega
=-1/2$, one obtains the critical point of spacetime as
\begin{eqnarray}
r_{+}^{c}&=&\notag 2.64432, {x^{c}}=0.656434,~~V^{c} =0.0253530\frac{4\pi ({r_{+}^{c})}^{3}}{3}=1.96363483\\
T_{eff}^{c} &=&\notag \frac{{0.2869}}{4\pi {r_{+}^{c}}}=0.00863395,~~P_{eff}=\frac{{%
0.10257633}}{8\pi ({r_{+}^{c})}^{2}}=0.000583686.
\end{eqnarray}
Substituting Eq. (\ref{3-9}) into Eqs. (\ref{3-2}) and (\ref{3-5}), we plot the isothermal curves of ${P_{eff}}-V$ in Fig. \ref{Fig7}. It shows that the value of $\omega $ has no effect on the isothermal curve of ${P_{eff}}-V$, while the curve of ${P_{eff}}-V$ is more sensitive to the parameter ${\beta _{{+}}}$.

It is shown in Fig. \ref{Fig7.1} that the isotherm of equation of state is in the range of $P_{eff}^{min} < {P_{eff}} < P_{eff}^{max}$, there are three possible values of $V$ which correspond to the same value of ${P_{eff}}$. Furthermore, for the range of $P_{eff}^{min} < {P_{eff}} < P_{eff}^{max}$, the slope of curve is positive, i.e., ${\left( {\frac{{\partial {P_{eff}}}}{{\partial V}}}\right)_{{T_{eff}},Q}} >0$, which does not meet the condition of stable equilibrium. Therefore, those states are impossible to
achieve as homogeneous systems. And the isotherm curves in the range of $P_{eff}^{min} < {P_{eff}} < P_{eff}^{max}$ are replaced with a straight line, so the areas enclosed by curve and straight are equivalent, that is the Maxwell's equal area law. With the increasing of the effective temperature, these two areas decrease. Until two areas tend to
zero, three intersection points of the straight line and the curve tend to one point, which is the critical point. The position of this critical point satisfies Eq. (\ref{4-2}).

\begin{figure}[tbp]
\centering
\subfigure[${Q} = 1,~\omega=-1/2$]{{\includegraphics[width=0.3\columnwidth,height=1.5in]{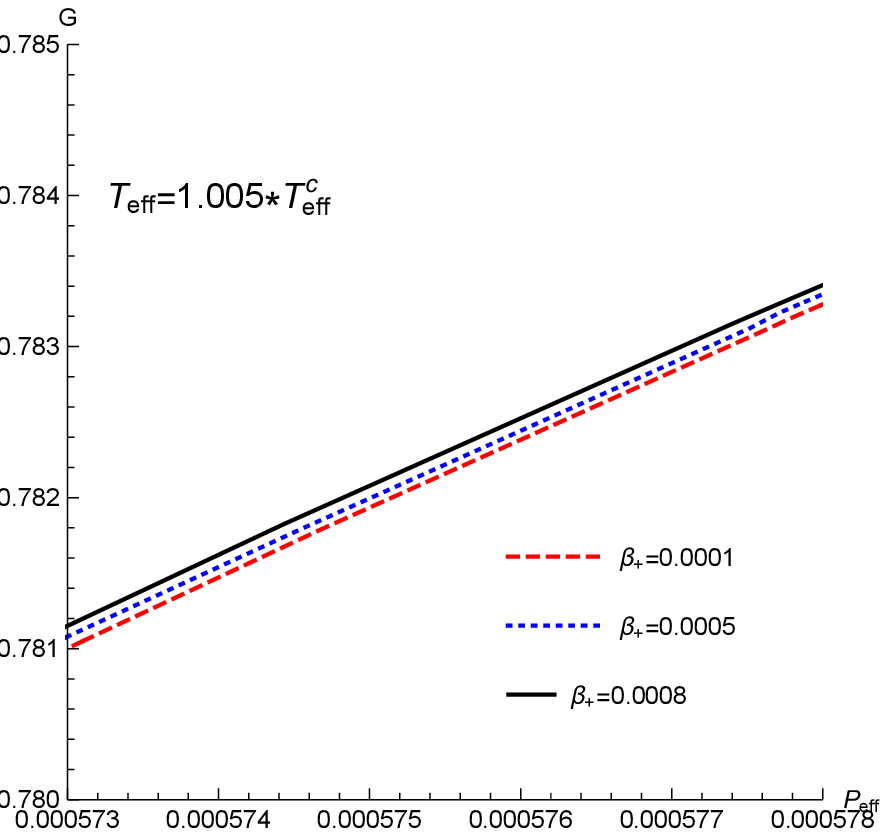}}\label{Fig8.1}}~~~~
\subfigure[${Q} = 1,~\omega=-1/2$]{{\includegraphics[width=0.3\columnwidth,height=1.5in]{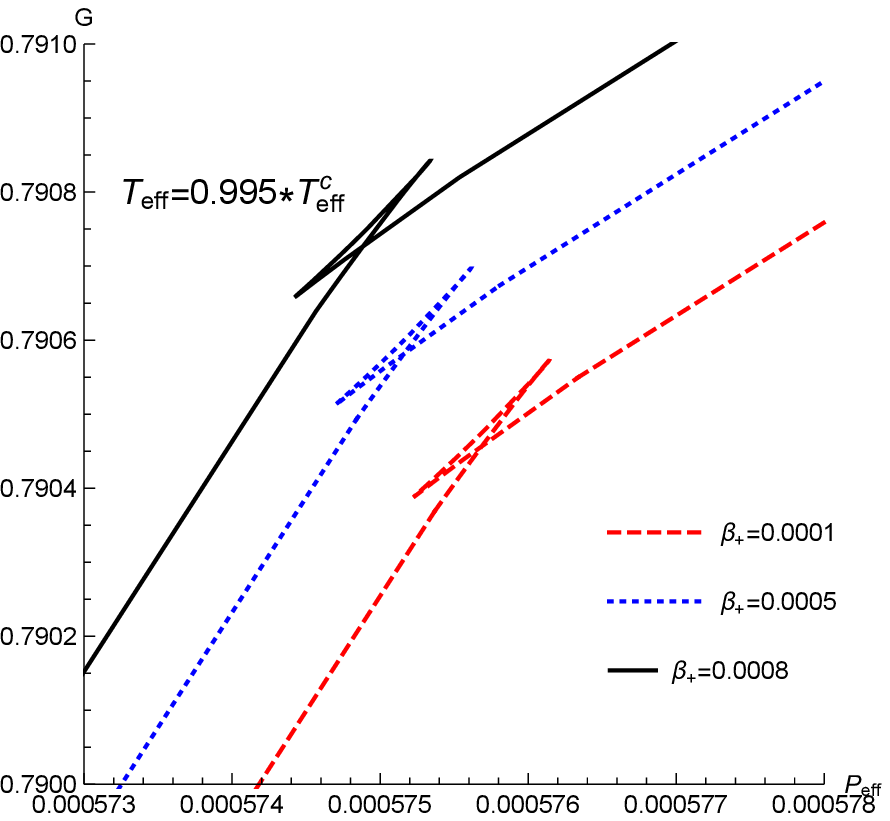}}\label{Fig8.2}}
\caption{(color online). ${G-P_{ eff}}$ curves with different values of ${\beta _{{+}}}$.}\label{Fig8}
\end{figure}

Substituting Eq. (\ref{3-9}) into Eqs. (\ref{3-5}) and (\ref{4-1}), we plot the isotherm curve of $G - {P_{eff}}$ in Fig. \ref{Fig8}.
The results show that when ${T_{eff}} > T_{eff}^c$, the isotherm curve $G - {P_{eff}}$ is monotonic. Namely, the RN-dSQ spacetime is in a single state, which corresponds to the gas phase of VdW system. When ${T_{eff}} < T_{eff}^c$, there exists a point of intersection in $G - {P_{eff}}$ curve,where the first-order phase transition occurs.

\begin{figure}[tbp]
\centering
\subfigure[$Q=1,~\omega=-1/2$]{{\includegraphics[width=0.3\columnwidth,height=1.5in]{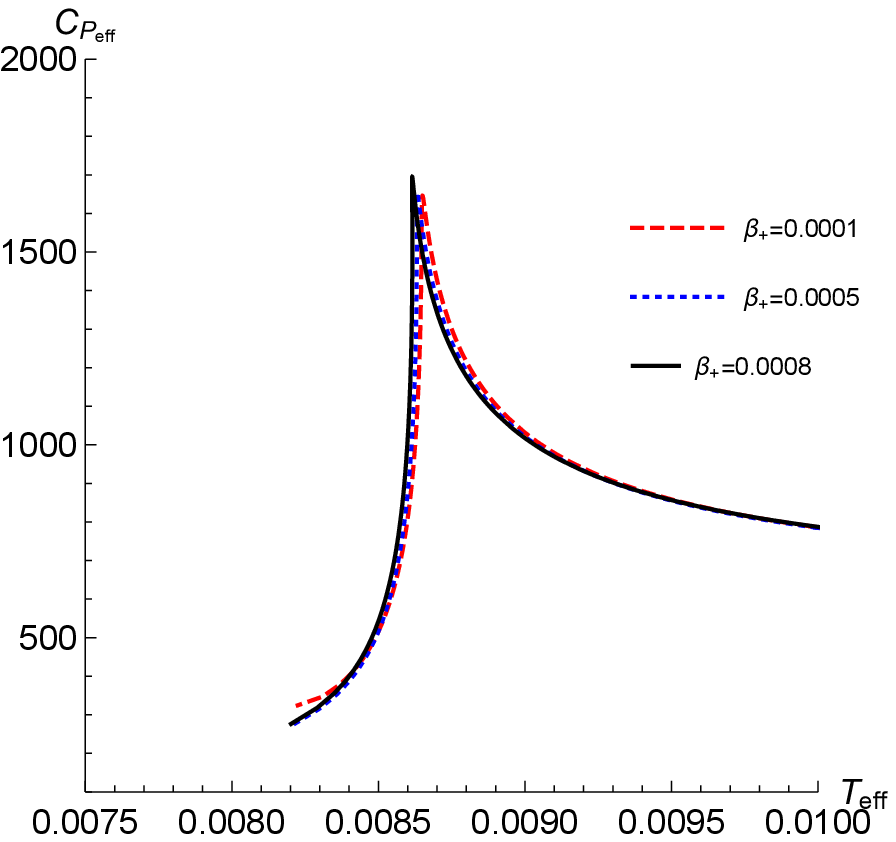}}\label{Fig9.1}}~~
\subfigure[$Q=1,~\omega=-1/2$]{{\includegraphics[width=0.3\columnwidth,height=1.5in]{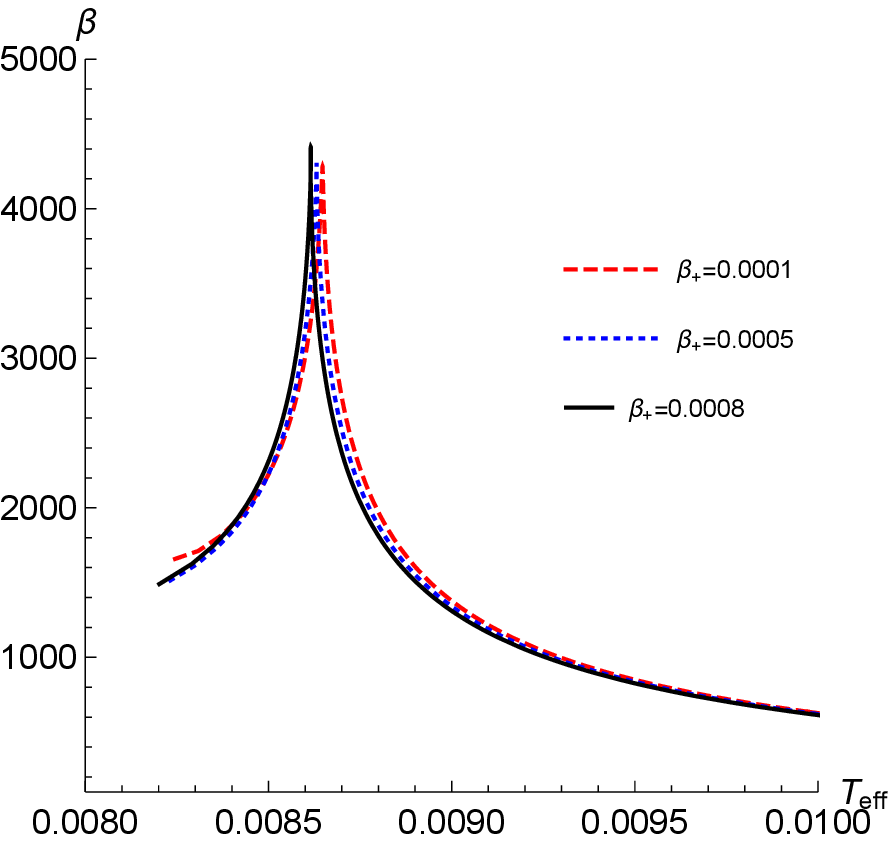}}\label{Fig9.2}}~~
\subfigure[$Q=1,~\omega=-1/2$]{{\includegraphics[width=0.3\columnwidth,height=1.5in]{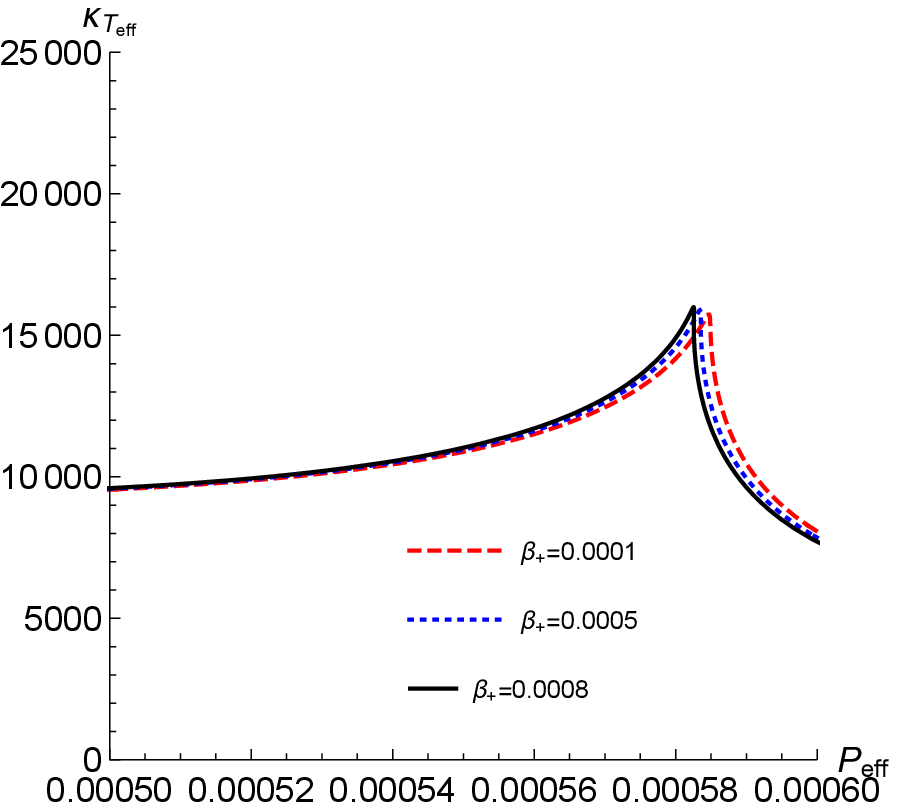}}\label{Fig9.3}}
\caption{(color online). Curves of ${C_{{P_{eff}}}} - {T_{eff}}$, $\beta - {T_{eff}}$, and ${\kappa _{{T_{eff}}}}- {P_{eff}}$ with different values of ${\beta _{{+}}}$. }\label{Fig9}
\end{figure}

The isobaric heat capacity of spacetime is
\begin{equation}
{C_{{P_{eff,Q}}}} = {T_{eff}}\frac{{\frac{{\partial S}}{{\partial x}%
}\frac{{\partial {P_{eff}}}}{{\partial {r_ + }}} - \frac{{\partial S}}{{%
\partial {r_ + }}}\frac{{\partial {P_{eff}}}}{{\partial x}}}}{{\frac{{%
\partial {T_{eff}}}}{{\partial x}}\frac{{\partial {P_{eff}}}}{{\partial {r_
+ }}} - \frac{{\partial {T_{eff}}}}{{\partial {r_ + }}}\frac{{\partial {%
P_{eff}}}}{{\partial x}}}}.\label{4-4}
\end{equation}
Substituting Eq. (\ref{3-11}) into Eq. (\ref{3-4}) and (\ref{4-4}), we have the isobaric curve of ${C_{%
{P_{eff}}}} - {T_{eff}}$ in Fig. \ref{Fig9.1}. The volume expansion coefficient reads
\begin{equation}
\beta = \frac{1}{V}\frac{{\frac{{\partial V}}{{\partial x}}%
\frac{{\partial {P_{eff}}}}{{\partial {r_ + }}} - \frac{{\partial V}}{{%
\partial {r_ + }}}\frac{{\partial {P_{eff}}}}{{\partial x}}}}{{\frac{{%
\partial {T_{eff}}}}{{\partial x}}\frac{{\partial {P_{eff}}}}{{\partial {r_
+ }}} - \frac{{\partial {T_{eff}}}}{{\partial {r_ + }}}\frac{{\partial {%
P_{eff}}}}{{\partial x}}}}.\label{4-5}
\end{equation}
Then substituting Eq. (\ref{3-11}) into Eqs. (\ref{3-4}) and (\ref{4-5}),  we plot the isobaric curve of
$\beta - {T_{eff}}$ in Fig. \ref{Fig9.2}. The isothermal compression coefficient becomes
\begin{equation}
{\kappa _{{T_{eff}}}} = \frac{1}{V}\frac{{\frac{{\partial V}}{{%
\partial x}}\frac{{\partial {T_{eff}}}}{{\partial {r_ + }}} - \frac{{%
\partial V}}{{\partial {r_ + }}}\frac{{\partial {T_{eff}}}}{{\partial x}}}}{{%
\frac{{\partial {T_{eff}}}}{{\partial x}}\frac{{\partial {P_{eff}}}}{{%
\partial {r_ + }}} - \frac{{\partial {T_{eff}}}}{{\partial {r_ + }}}\frac{{%
\partial {P_{eff}}}}{{\partial x}}}}.\label{4-6}
\end{equation}
Substituting Eq. (\ref{3-9}) into Eqs. (\ref{3-5}) and (\ref{4-6}), we plot the isothermal curve of ${%
\kappa _{{T_{eff}}}} - {P_{eff}}$ in Fig. \ref{Fig9.3}. Fig. \ref{Fig9} shows that RN-dSQ spacetime presents typical continuous phase transition characteristics, i.e., it possesses the phase transition characteristics of VdW system or AdS black holes. These curves also reflect the influence of parameter ${\beta _ + }$ on the critical point.

\begin{figure}[tbp]
\centering
{\includegraphics[width=0.3\columnwidth,height=1.5in]{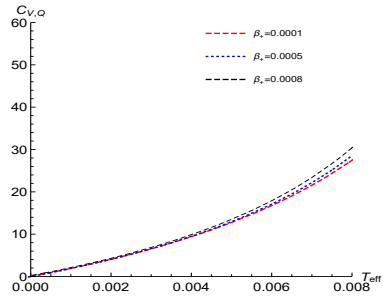}}
\par
\caption{(color online). ${C_{V,Q}} - {T_{eff}}$ with different values of ${\beta _{{+}}}$ for $Q=1$, $\omega=1/2$.}\label{Fig10}
\end{figure}

The isochoric heat capacity of spacetime is
\begin{equation}
{C_{V,Q}} = {T_{eff}}{\left( {\frac{{\partial S}}{{\partial {T_{eff}}}}}
\right)_{V,Q}} = {T_{eff}}\frac{{\frac{{\partial S}}{{\partial x}}\frac{{%
\partial V}}{{\partial {r_ + }}} - \frac{{\partial S}}{{\partial {r_ + }}}%
\frac{{\partial V}}{{\partial x}}}}{{\frac{{\partial {T_{eff}}}}{{\partial x}%
}\frac{{\partial V}}{{\partial {r_ + }}} - \frac{{\partial {T_{eff}}}}{{%
\partial {r_ + }}}\frac{{\partial V}}{{\partial x}}}}.\label{4-7}
\end{equation}
Substituting Eq. (\ref{3-2}) into Eqs. (\ref{3-4}) and (\ref{4-7}), we present the isochoric curve of ${%
C_{V,Q}} - {T_{eff}}$ in Fig. \ref{Fig10}.
It can be seen from Fig. \ref{Fig10} that, the isochoric heat capacity of spacetime doesn't vanish. That is different with AdS black holes, whose isochoric heat capacities are zero \cite{wei-2019, wei-2020,zou-2020,wei-202011}. For any thermodynamic systems, the isochoric heat capacity doesn't vanish in general, so our conclusion is more universal. Therefore, when considering the interaction of two horizons for the RN-dSQ spacetime, the effective thermodynamic quantities and corresponding equations are more closer to the characteristics of an ordinary thermodynamic system. Thus this thermodynamic system described by the effective thermodynamic quantities is more universal.

According to Ehrenfest's classification of phase transitions, if the Gibbs functions of two phases and their first-order partial derivatives are continuous, but their second-order partial derivatives are of sudden changes, the phase transition is called the second-order one. From the above analysis, we know that for the lower effective
temperature (${T_{eff}} < T_{eff}^c$), the entropy and volume will undergo sudden changes, and the spacetime undergoes a first-order phase transition. As ${T_{eff}} = T_{eff}^c$, the entropy and volume are continuous, while the isobaric heat capacity, isobaric expansion coefficient, and isothermal compression coefficient both have a peak at the critical point shown in Fig. \ref{Fig9}.

\section{Entropic force between two horizons}
\label{5}
\begin{figure}[tbp]
\centering
\subfigure[${l^{2}}=37.386,~{\beta _{{+}}}=0.0005$]{
{\includegraphics[width=0.3\columnwidth,height=1.5in]{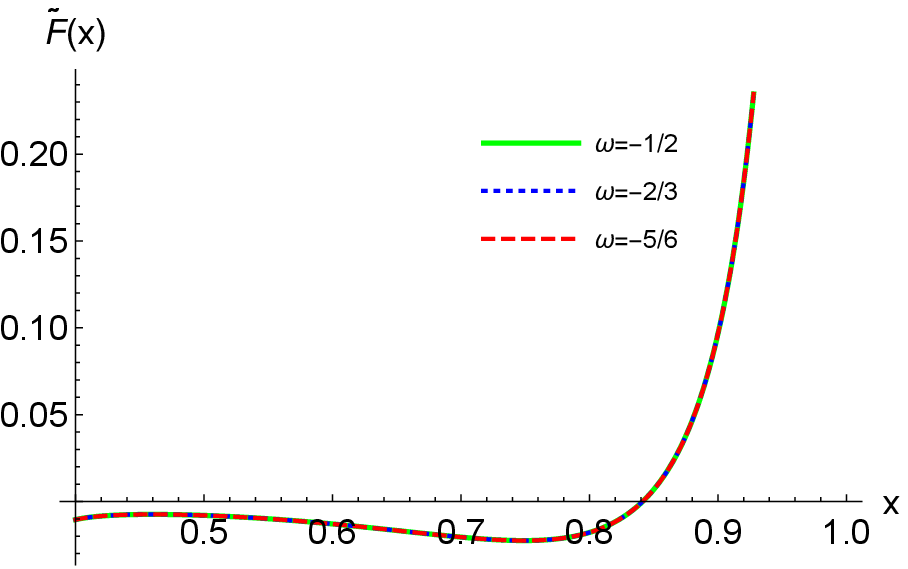}}\label{Fig11.1}}
\subfigure[$\omega=-2/3,~{l^{2}}=37.386$]{
{\includegraphics[width=0.3\columnwidth,height=1.5in]{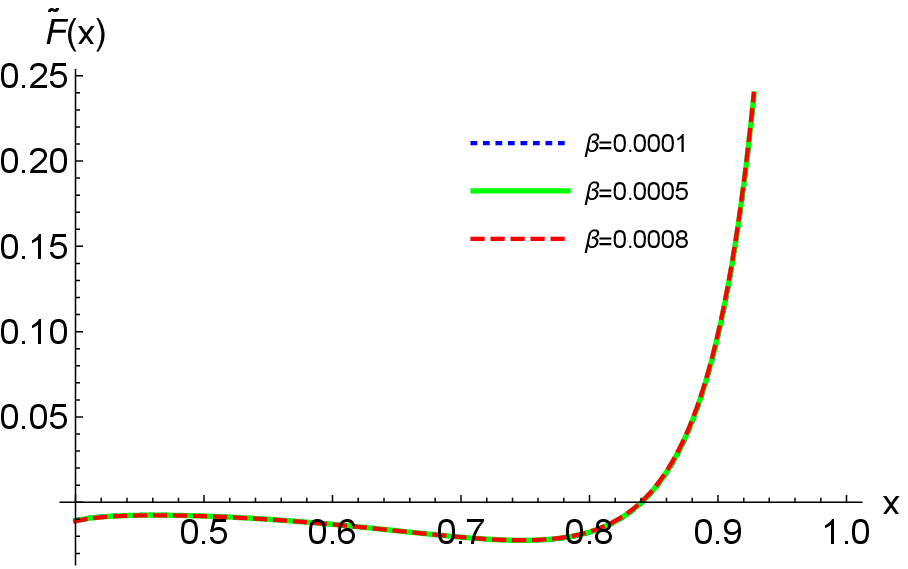}}\label{Fig11.2}}
\subfigure[$\omega=-2/3,~{\beta _{{+}}}=0.0005$]{
{\includegraphics[width=0.3\columnwidth,height=1.5in]{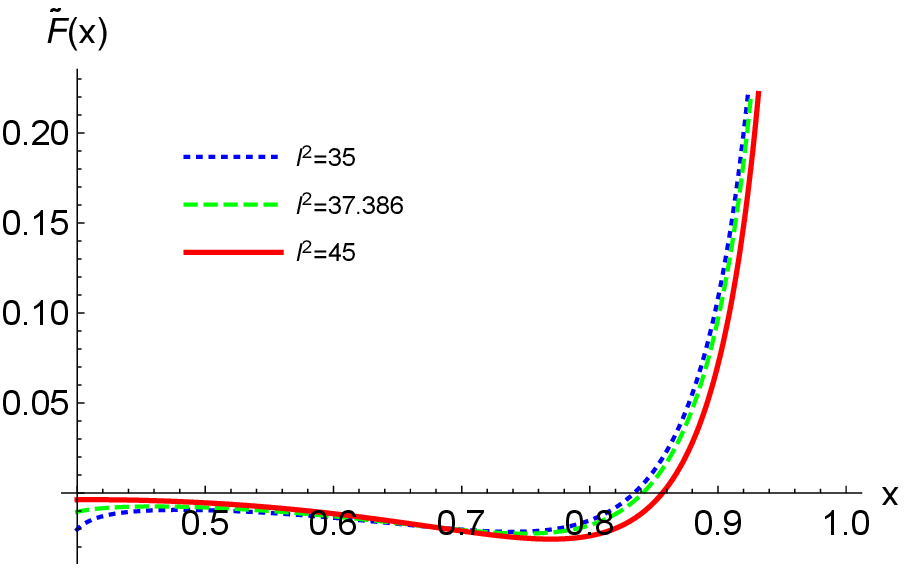}}\label{Fig11.3}}
\caption{(color online). Curves of $\tilde F(x) - x$ with $Q=1$.}\label{Fig11}
\end{figure}

The recent exploration of the interactive force between microscopic particles inside black holes has attracted widespread attention \cite{wei-2019,Dinsmore-2020, Clifford-20202020,wei-2021,Suvankar-2021}. In order to understand the microstructure of RN-dSQ spacetime more clearly, we will check out the behavior of entropic force between two horizons in this section. The entropic force in a thermodynamic system can be defined as follows \cite{zhao-2021,zhang-2019,Verlinde-2011}
\begin{equation}
F=-T\frac{{\partial S}}{{\partial r}},\label{5-1}
\end{equation}
where $T$ is the temperature and $r$ is the radius of system. In this part we pay attention on the entropic force between the black hole and cosmological horizons. According to Eq. (\ref{3-3}), the interactive entropy caused by two horizons reads
\begin{equation}
\tilde S = \pi r_c^2f(x).
\end{equation}
Based on Eq. (\ref{5-1}), we can introduce the corresponding entropic force between two horizons in the RN-dSQ spacetime
\begin{equation}
\tilde F = {T_{eff}}{\left( {\frac{{\partial \tilde S}}{{\partial r}}} \right)_{{T_{eff}}}},\label{5-3}
\end{equation}
where $T_{eff}$ stands for the effective temperature of this system and $r\equiv{r_c}-{r_ + }$. From the above equation, we have
\begin{equation}
\tilde F(x) = {T_{eff}}{x^2}\frac{{{{\left( {\frac{{\partial \tilde S}}{{\partial {r_ + }}}} \right)}_x}{{\left( {\frac{{\partial {T_{eff}}}}{{\partial x}}} \right)}_{{r_ + }}} - {{\left( {\frac{{\partial \tilde S}}{{\partial x}}} \right)}_{{r_ + }}}{{\left( {\frac{{\partial {T_{eff}}}}{{\partial {r_ + }}}} \right)}_x}}}{{x(1 - x){{\left( {\frac{{\partial {T_{eff}}}}{{\partial x}}} \right)}_{{r_ + }}} + {r_ + }{{\left( {\frac{{\partial {T_{eff}}}}{{\partial {r_ + }}}} \right)}_x}}}. \label{5-4}
\end{equation}
Then substituting Eq. (\ref{3-4}) into Eqs. (\ref{4-2}) and (\ref{5-4}), we can get the $\tilde F(x) - x$ curves as shown in Fig. \ref{Fig11}.

It is obviously that the entropic force between two horizons depends on the cosmological constant ${l^2}$, while merely independent of other parameters ${\beta _{{+}}}$ and $\omega$. For the far distance between two horizons (i.e., the small value of $x$), the entropy force is negative, which indicates there exists an attractive force between two horizons to make them more closer. Up to the certain distance (i.e., the certain value of $x$), the entropic force is zero. Continuously decreasing the distance, the interactive entropy force between two horizons becomes positive, and it will approach to infinity when two horizons coincide. In the case of $x=1$, the infinite entropic force prompts them to accelerated separate.

\section{Conclusion and discussion}
\label{6}

In this work, we introduced the interaction between the black hole and cosmological horizons and presented the effective thermodynamic quantities of the RN-dSQ spacetime. The related thermodynamical properties of system were also investigated. Through the analysis in Section IV, we know that the RN-dSQ spacetime also have the continuous phase transitions and first-order ones, that is similar to VdW system or AdS black holes. The results showed that the effective temperature and phase transition point are independent of the RN-dSQ spacetime quintessence dark energy barotropic index $\omega$. The existence of the black hole and cosmological horizons are determined by the cosmological constant and the normalization factor associated with the quintessence density of dark energy $\alpha$ when other parameters are given. Based on the effective quantities of system, we found that the isochoric heat capacity is non-zero, and it increases with the effective temperature and quintessence field ${\beta _{\mathrm{\ +}}}$. This conclusion is similar to ordinary thermodynamic systems, so our results have more general physical meanings.

From the curve of $\tilde F(x) - x$, the characteristic of the change of entropic force with $x$ is: when $x \to 1$, the entropic force tends to infinity, which indicates that the infinite entropic force leads to accelerated separation between two horizons. At the same time, two horizons will be accelerating separattion under the infinity entropic force, and the energy $M$ of system will decrease. Continuously decreasing the value of $x$, the entropic force will become negative and prevent the separation of two horizons. On the other hand, from the curve ${T_{eff}} - x$ we found that for the low effective temperature (i.e., the large value of $x$), the entropic force between the two horizons promotes the separation of two
horizons. While for the high effective temperature (i.e., the small value of $x$), the entropic force prevents the separation of the two horizons. These conclusions will provide a theoretical foundation for studying the evolution of spacetime.

In addition, comparing the entropic force between two horizons with the Lennard-Jones force between two particles, we found that the curves obtained by completely different methods for different systems are so surprisingly similar. The Lennard-Jones force of two particles is inherently related to the entropic force between two horizons. To further study the thermodynamic properties of black holes, it is necessary to explore the microstructure of system. We had shown the interactive force between two horizons, which reflected the interactive force between microscopic particles on the curved surfaces (the black hole and cosmological horizons) in dS spacetime. The method can be extended to study the interaction of microscopic particles inside black holes, which will provides a new way to explore the interaction between black hole molecules.

\section*{Acknowledgments}

We thank Prof. M. S. Ma and Prof. Z. H. Zhu for useful discussions.

This work was supported by the National Natural Science Foundation of China (Grant Nos. 12075143, 11971277),
the Scientific and Technological Innovation Programs of Higher Education Institutions of Shanxi Province, China (Grant No. 2020L0471, No. 2020L0472) and the Science Technology Plan Project of Datong City, China (Grant No. 2020153).

\end{document}